\documentclass[aps,prb,twocolumn,showpacs,superscriptaddress]{revtex4}

\usepackage{epsfig}
\usepackage{psfrag}
\usepackage{graphicx}
\usepackage{amsfonts}
\usepackage[figuresright]{rotating}
\usepackage{amssymb}
\usepackage{amsmath}
\usepackage{subfigure}
\usepackage{multirow}
\usepackage{tabularx}
\usepackage{bm}
\usepackage{array}
\usepackage[usenames]{color}

\def\posv2{r} 
\def\latvec{\mathbf{l}} 
\def\basisvec{\mathbf{b}} 
\def\uvec{\mathbf{u}} 
\def\momvec{\mathbf{k}} 
\def\momv2{k} 
\def\magvec{\mathbf{\kappa}} 
\def\cellvec{\mathbf{R}} 
\def\toppol{\mathbf{R}_T} 
\def\zone{z_1} 
\def\ztwo{z_2} 
\def\det{\textrm{det}} 
\def\phase{\phi} 

\newcommand{\rv}{{\bm r}}
\newcommand{\ev}{{\bm e}}
\newcommand{\bv}{{\bm b}}

\newcommand{\Gv}{\bm G}

\newcommand{\lv}{\mathbf{l}}
\newcommand{\uv}{{\bm u}}
\newcommand{\tv}{{\bm t}}
\newcommand{\fv}{{\bm f}}

\newcommand{\kv}{{\bm k}}

\newcommand{\av}{{\bm a}}

\newcommand{\Qv}{{\mathbf Q}}
\newcommand{\Cv}{{\mathbf C}}
\newcommand{\Dv}{{\mathbf D}}

\newcommand{\Rv}{{\bm R}}

\newcommand{\Av}{{\mathbf A}}

\newcommand{\st}{\varepsilon}
\newcommand{\stm}{\pmb{\st}}

\newcommand{\trem}[1]{{\color{green}{}}}

\begin{document}
\title{Mechanical Weyl Modes in Topological Maxwell Lattices}

\author{D. Zeb Rocklin}
\affiliation{Department of Physics, University of Michigan, 450
Church St. Ann Arbor, MI 48109, USA}

\author{Bryan Gin--ge Chen}
\altaffiliation{Current address: Department of Physics,
University of Massachusetts, Amherst, MA 01002, USA}
\affiliation{Instituut-Lorentz, Universiteit Leiden, 2300 RA
Leiden, The Netherlands}

\author{Martin Falk}
\affiliation{Department of Physics, Massachusetts Institute of
Technology, 77 Massachusetts Avenue Cambridge, MA 02139-4307,
USA}

\author{Vincenzo Vitelli}
\affiliation{Instituut-Lorentz, Universiteit Leiden, 2300 RA
Leiden, The Netherlands}

\author{T. C. Lubensky}
\affiliation{Department of Physics and Astronomy, University of
Pennsylvania, Philadelphia, PA 19104, USA }

\date{\today}

\begin{abstract}

Topological mechanical structures exhibit robust properties
protected by topological invariants. In this letter, we study a
family of deformed square lattices that display topologically
protected zero-energy {\it bulk} modes analogous to the
massless fermion modes of Weyl semimetals. Our findings apply
to sufficiently complex lattices satisfying the Maxwell
criterion of equal numbers of constraints and degrees of
freedom. We demonstrate that such systems exhibit pairs of
oppositely charged Weyl points, corresponding to zero-frequency
bulk modes, that can appear at the origin of the Brillouin zone
and move away to the zone edge (or return to the origin) where
they annihilate. We prove that the existence of these Weyl
points leads to a wavenumber-dependent count of topological
mechanical states at free surfaces and domain walls.

\end{abstract}

 \maketitle

\date{\today}

Topological properties of the energy operator and associated
functions in wavevector (momentum) space can determine
important properties of physical systems
\cite{Nakahara2003,volovik03,Volovik2007}. In quantum condensed
matter systems, topological invariants guarantee the existence
and robustness of electronic states at free surfaces and domain
walls in polyacetylene \cite{ssh,jackiw76}, quantum Hall
systems \cite{halperin82,haldane88} and topological insulators
\cite{km05b,bhz06,mb07,fkm07,HasanKane2010,QiZhang2011} whose
bulk electronic spectra are fully gapped (i.e. conduction and
valence bands separated by a gap at all wavenumbers). More
recently topological \emph{phononic} and \emph{photonic} states
have been identified in suitably engineered classical materials
as well,
\cite{Prodan09,KaneLub2014,PauloseVit2015,PauloseVit2015-b,ChenVit2014,VitelliChe2014,Chen2015,Xiao2015,Po2014,Yang2015,Nash2015,Wang2015,Wang2015a,Susstrunk2015,Kariyado2015,Peano2015,Mousavi2015,Khanikaev2015}
provided that the band structure of the corresponding wave-like
excitations has nontrivial topology.

A special class of topological mechanical states occurs in
\emph{Maxwell lattices}, periodic structures in which the
number of constraints equals the number of degrees of freedom
in each unit cell~\cite{LubenskySun2015}. In these mechanical
frames, zero-energy modes and states of self stress (SSS) are
the analogs of particles and holes in electronic topological
materials \cite{KaneLub2014}. A zero energy (frequency) mode is
the linearization of a {\em mechanism}, a motion of the system
in which no elastic components are stretched.
States of self stress on the other hand guide the focusing of
applied stress and can be exploited to selectively pattern
buckling or failure~\cite{PauloseVit2015-b}.
Such mechanical states can be topologically protected in
Maxwell lattices, such as the distorted kagome lattices of
Ref.~\onlinecite{KaneLub2014}, in which no zero modes exist in
the {\it bulk} phonon spectra (except those required by
translational invariance at wavevector $\kv=0$). These lattices
are the analog of a fully gapped electronic material.
 They are
characterized by a topological polarization equal to a lattice
vector $\Rv_T$  that, along with a local polarization $\Rv_L$,
determines the number of zero modes {\it localized} at free
surfaces, interior domain walls separating different
polarizations, and dislocations~\cite{PauloseVit2015}. Because
$\Rv_T$ only changes upon closing the bulk phonon gap, these
modes are robust against disorder or imperfections.

\begin{widetext}

\begin{figure}
\subfigure{\includegraphics[width=.38\textwidth,height=.4\textwidth]{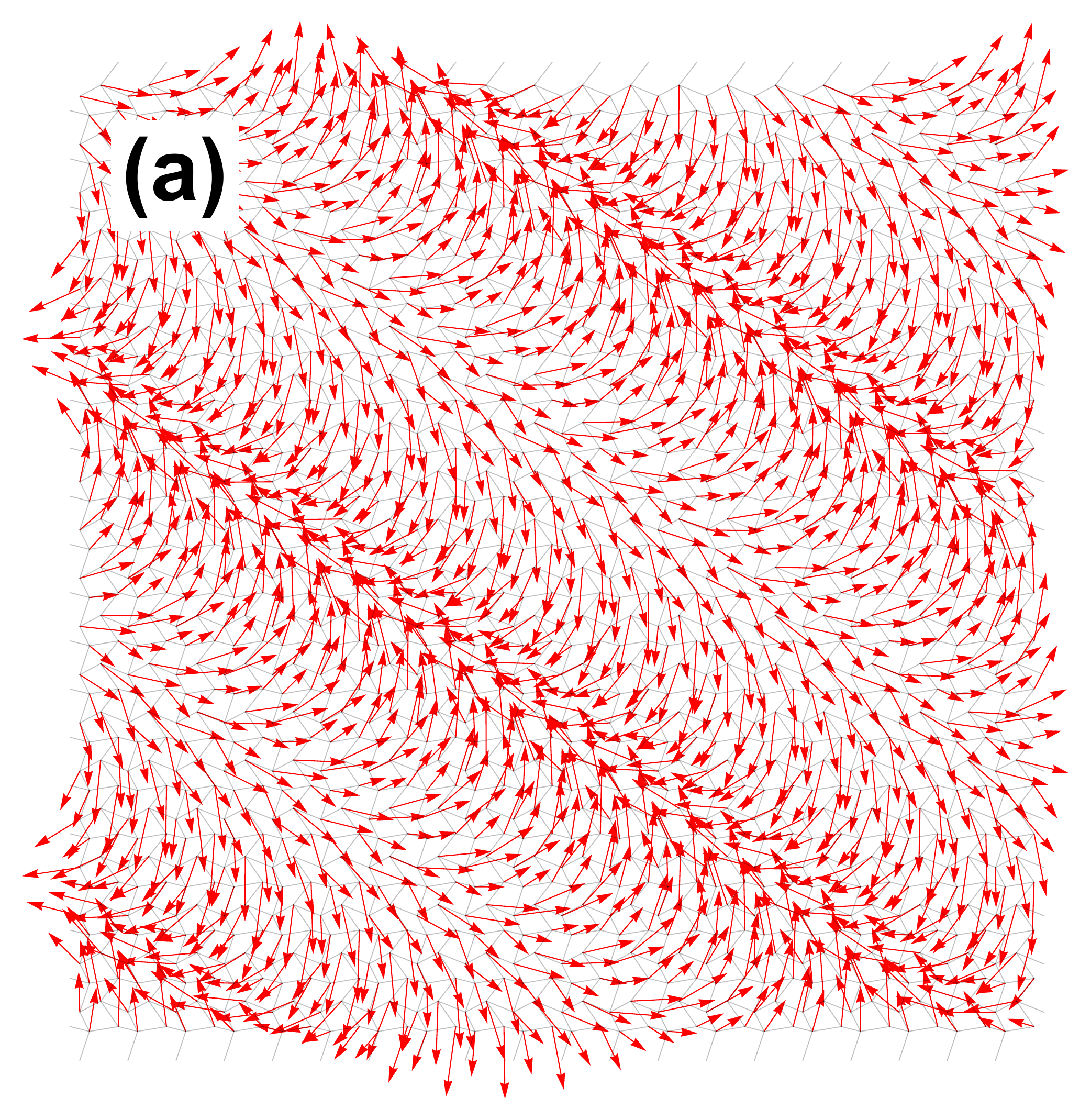}
\llap{
\includegraphics[width=.15\textwidth]{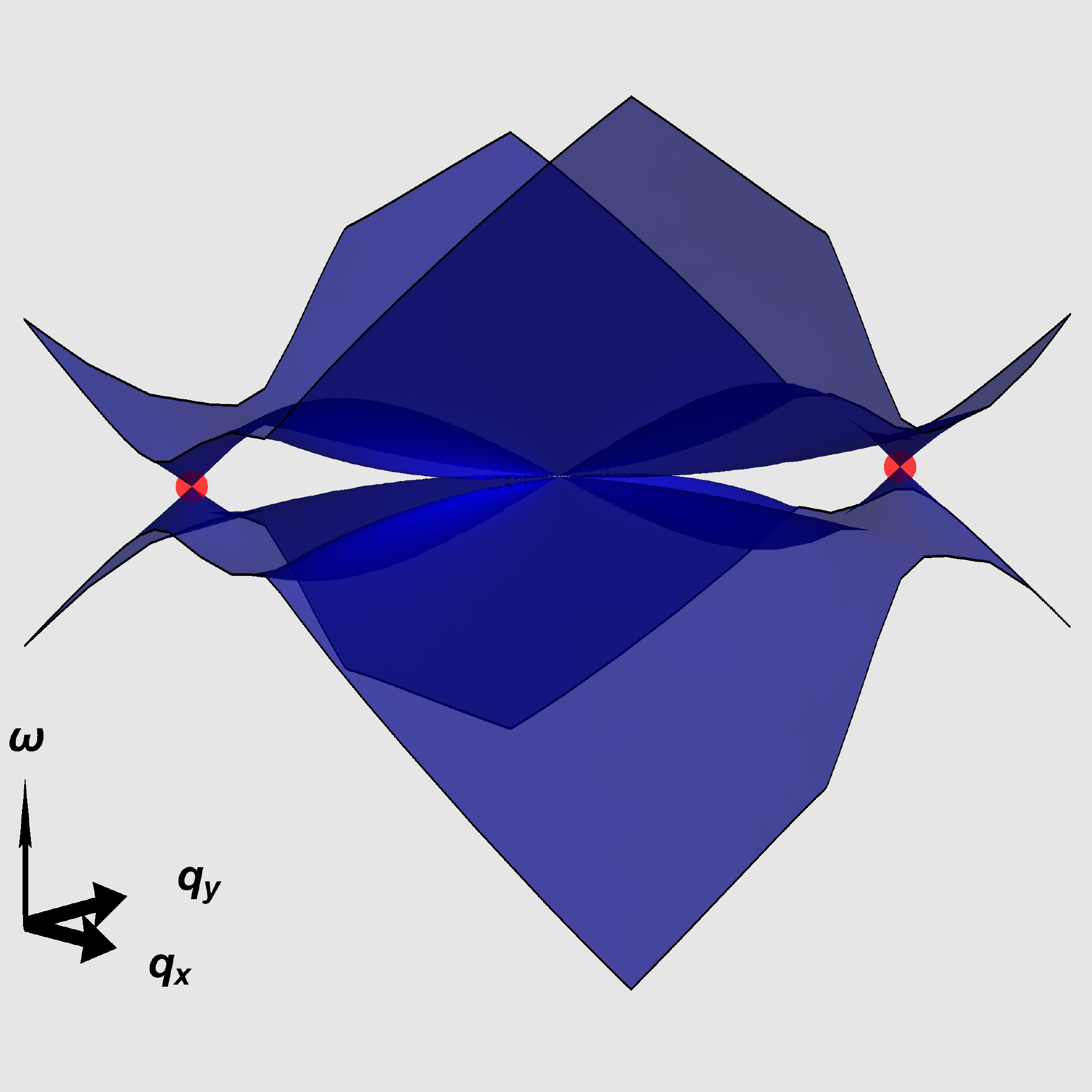}
}}
\subfigure{\includegraphics[width=.38\textwidth,height=.4\textwidth]{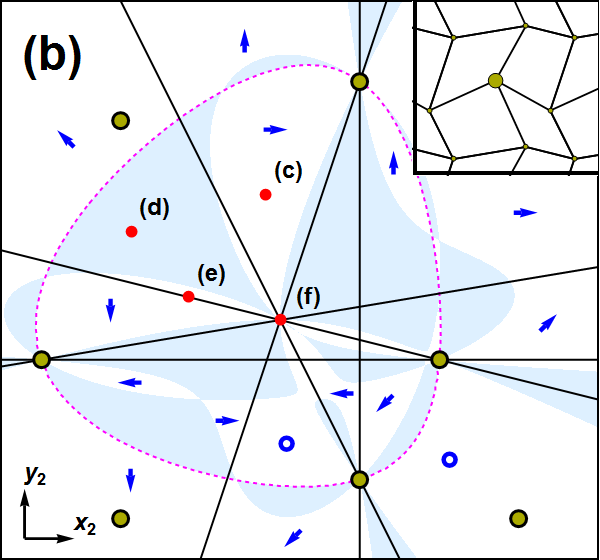}}
\subfigure{\includegraphics[width=.2\textwidth]{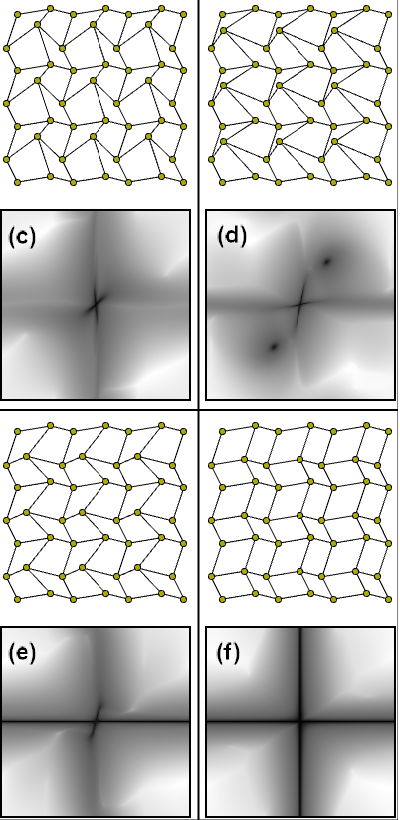}}
\caption{
(a) Deformed square lattices can have sinusoidal bulk zero modes (red arrows) corresponding to Weyl points where two bands touch in the phonon dispersion (inset).
(b) The phase diagram of a deformed square lattice with the positions
of three sites fixed and the position of the remaining site given by
the position $(x_2, y_2)$, shown as an enlarged site in the inset.
(c)-(f) Lattices (top) with phonon dispersions (bottom) with dark areas indicating low-energy modes in the Brillouin Zone.
In (b),
white areas such as (c) lack Weyl points and are marked with their
a blue arrow indicating their topological polarization. Blue-shaded areas such as (d) correspond to Weyl lattices.
Open boundaries between white and blue regions indicate where Weyl points
emerge at $\kv=(0,0)$ while the pink dashed boundary indicate where
they annihilate at $\kv=(\pi,\pi)$. Lattices on the Special Lines, such as (e) lie
between topological polarizations and possess lines of zero
modes along $k_{x(y)} =0$, while at the Special Point, (f),
there are two zero modes along each of $k_{x(y)}=0$. }
\label{fig:phase-dia}
\end{figure}

\end{widetext}

In this paper we demonstrate how to create topologically
protected zero modes and states of self-stress that extend
throughout a sample. These enable the topological design of
bulk soft deformation and material failure in a generic class
of mechanical structures. As prototypes, we study the distorted
square lattices of masses and springs shown in
Fig.~\ref{fig:phase-dia}, and we show that they have phases
that are mechanical analogs of Weyl semi-metals
\cite{WanVish2011,BurkovBal2011a,BurkovBal2011,LiuVan2014,XuHa2015}.
In the latter materials, the valence and conduction bands touch
at isolated points in the Brillouin zone (BZ), with the
equivalent \emph{phonon} dispersion for the mechanical lattice
shown in Fig.~\ref{fig:phase-dia}(a). Points at which two or
four bands touch are usually called Weyl and Dirac points,
respectively. These points, which are essentially wavenumber
vortices in 2D and hedgehogs in 3D, are characterized by a
winding or Chern number \cite{Nakahara2003,Volovik2007} and are
topologically protected in that they can disappear only if
points of opposite sign meet and annihilate or if
symmetry-changing terms are introduced into the Hamiltonian.
Weyl semi-metals exhibit lines of surface states at the Fermi
level that terminate at the projection of the Weyl points onto
the surface BZ. Weyl points have also been predicted
\cite{LuJoan2013} and observed \cite{LuSol2015} in photonic
crystals and certain mechanical systems with pinning
constraints that gap out translations \cite{Po2014}. In
contrast, our examples consist of ordinary spring networks
which suggest that the Weyl phenomenon is in fact generic to
Maxwell lattices of sufficient complexity.

The full phase space of our distorted square lattices is
six-dimensional; to keep our discussion simple, we fix three of
the four sites in the unit cell and vary the equilibrium
coordinates $(x_2,y_2)$ of one of the sites. The resultant 2D
phase diagram, shown in Fig.~\ref{fig:phase-dia}(b), exhibits a
rich phenomenology: (1) a special point (SP) at the origin with
two orthogonal lines of zero modes in its spectrum, (2) special
lines (SLs) along which the spectrum exhibits a single line of
zero modes, (3) finite regions in which the spectrum is fully
gapped and characterized by topological polarizations $\Rv_T$,
and (4) finite regions whose spectrum contains Weyl points.  In
(4), pairs of oppositely charged Weyl points  corresponding to
zero-frequency {\em bulk} modes appear at the origin of the BZ
and then move away to the zone edge or back to the origin where
they annihilate. The existence of Weyl points has significant
consequences for response at the boundaries, leading to a
wavenumber-dependent count of boundary modes and SSS.

Lattices of periodically repeated unit cells in $d$ dimensions
with $n$ sites (nodes) and $n_B$ bonds per unit cell under
periodic boundary conditions (PBCs) are characterized
\cite{SunLub2012,LubenskySun2015} by an $n_B \times d n$
compatibility matrix $\Cv(\kv)$ for each wavevector $\kv$ in
the BZ relating the $d n$-dimensional vector $\uv(\kv)$ of site
displacements to the $n_B$-dimensional vector $\ev(\kv)$ of
bond extensions via $\Cv(\kv) \uv = \ev(\kv)$ and by the $dn
\times n_B$ equilibrium matrix $\Qv (\kv) = \Cv^\dag (\kv)$
relating forces $\fv(\kv)$ to bond tensions $\tv(\kv)$ via
$\Qv(\kv) \tv(\kv) = \fv(\kv)$. The dynamical matrix (for
systems with unit masses and spring constants) is $\Dv =
\Qv(\kv) \Cv(\kv)$. Vectors $\uv(\kv)$ in the null space of
$\Cv(\kv)$ do not stretch bonds and, therefore, correspond to
zero modes. Vectors $\tv(\kv)$ in the null space of $\Qv(\kv)$
describe states with tensions but without net forces and thus
correspond to SSSs~\cite{Calladine1978}. The number of zero
modes $n_0(\kv)$ and SSSs $n_s(\kv)$ at each $\kv$ are related
by the Calladine-Maxwell index theorem \cite{Calladine1978}
\begin{equation}
\nu(\kv) \equiv n_0(\kv) - n_s(\kv) = dn - n_B .
\end{equation}
Lattices, like the square and kagome lattices in two
dimensions, are \emph{Maxwell lattices} that have the special
property that $n_B = 2n = dn$ and, as a result, $n_0(\kv) =
n_s(\kv)$ for every $\kv$. This relation states that for every
$\kv$ there is one zero mode for each SSS and vice versa. Thus
in a gapped system, there are no SSSs for any $\kv \neq 0$, and
at $\kv = 0$, translational invariance requires $n_0(\kv) \geq
d$ and thus $n_s(\kv) \geq d$. A Weyl point by definition is a
$\kv$ at which there is a zero mode, and there is necessarily
an SSS that goes with it. The lines of zero modes occurring
along the SLs in the phase diagram also have associated lines
of self-stress in real space.

We now turn to zero modes on free surfaces. Cutting a lattice
under PBCs along a direction perpendicular to one of its
reciprocal lattice vectors $\Gv$ creates a finite-width strip
with two free surfaces aligned along the ``parallel'' direction
perpendicular to $\Gv$. The cut removes $\Delta n_B$ bonds and
$\Delta n$ sites for each unit cell along its length or
equivalently for each wavenumber $-\pi/a_{||}< q< \pi/a_{||}$
along the cut, where $a_{||}$ is the length of the surface
cell. The index theorem relating the total number of zero modes
$N_0(q,\Gv)$ to the total number of SSSs $N_s(q,\Gv)$ at each
$q$ is
\begin{equation}
N_0(q,\Gv)-N_s(q,\Gv) = - d \Delta n + \Delta n_B .
\label{eq:gaps1}
\end{equation}
Bulk modes in the spectrum are described by the same $\Cv(\kv)$
as the uncut sample but with a different set of quantized
wavenumbers $p$ parallel to $\Gv$. If a bulk mode is gapped in
the periodic spectrum, it remains gapped in the strip without
an associated state of self-stress. Thus if the bulk modes are
gapped at a given $q$ under PBCs, their contribution to the
right-hand side of Eq.~(\ref{eq:gaps1}) will be zero, implying
that only surface modes contribute to Eq.~(\ref{eq:gaps1}). In
addition, cutting the sample will not introduce additional
SSSs. The result is that Eq.~(\ref{eq:gaps1}) becomes an
equation for the total number of zero surface modes on both
surfaces:
\begin{equation}
n_0^{ST} (q,\Gv) = \Delta n_B - d \Delta n .
\end{equation}
This is a global relation that applies to every $q$ in the
surface BZ at which the bulk spectrum is gapped.

\begin{figure}
\includegraphics[width=0.45\textwidth]{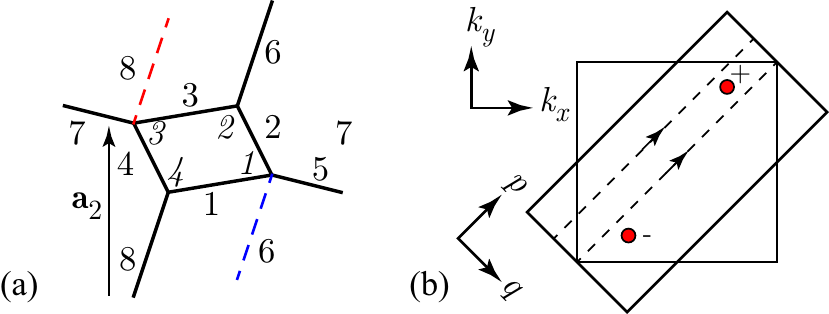}
\caption{(a) Different versions of the unit cell with four sites $1-4$ labeled in italic script.
The cell consisting of bonds, labeled $1-8$ in roman script and drawn in full black, is the
symmetric unit cell. A unit cell associated with a lower (an upper) surface parallel to the
$x$-axis is constructed by moving moving bond $8$
through $\av_2$ (bond $6$ through $-\av_2$) to the dashed red (blue)
line to yield $\Rv_L=-\av_2$ ($\Rv_L=\av_2$).
(b) depicts the standard BZ with two Weyl points and the BZ
dual to a surface-compatible unit cell oriented at 45$^\circ$. The
component of $\kv$ along the surface is $q$ and that parallel
to $\Gv$ is $p$. It also shows two paths, one on each side of
the projected position $q_w^+$ the ``$+$'' Weyl point at $\kv_w^+$.}
\label{fig:weylpath}
\end{figure}

In the bulk, our lattice is naturally described by a symmetric
unit cell as depicted in Fig.~\ref{fig:weylpath}(a). We
associate with each lattice site a ``charge'' $+2$ and with the
center of each bond a charge $-1$~\cite{KaneLub2014}. The total
dipole moment of this cell is zero.
The number of zero modes for a surface perpendicular to $\Gv$
can be calculated from the compatibility matrix $\Cv(\kv,\Gv)
\equiv \Cv(q,p,\Gv)$ of another unit cell, one that is
compatible with the surface $\Gv$ (Fig.~\ref{fig:weylpath}(b)).
Its components are related to those of the compatibility matrix
, $\tilde{\Cv}(\kv)$, of the symmetric unit cell via
$C_{\beta,\sigma i} = e^{i \kv\cdot \Delta \rv_\beta}
\tilde{C}_{\beta,\sigma i} e^{-i \kv\cdot \Delta \rv_\sigma}$,
where $\beta = 1, \cdots, n_B$ labels bonds, $\sigma = 1,
\cdots, n$ labels sites, and $i = x,y$ and $\Delta \rv_\beta$
and $\Delta \rv_\sigma$ are, respectively, the displacements of
bond $\beta$ and site $\sigma$ necessary to convert the
symmetric unit cell to the surface-compatible one. Both
$\rv_\beta$ and $\rv_\sigma$ are necessarily Bravais lattice
vectors. Thus,
\begin{equation}
\det \, \Cv(\kv,\Gv) = \exp(- i \kv \cdot \Rv_L)\det \,\tilde{\Cv} (\kv) ,
\end{equation}
where $\Rv_L = 2 \sum_\sigma \Delta \rv_\sigma - \sum_\beta
\Delta \rv_\beta$. $\det \,\Cv(\kv, \Gv)$ is invariant under $p
\rightarrow p + G$, where $G = |\Gv|$, and is a polynomial in
$z=\exp (i 2 \pi p /G)$ with no negative powers of $z$ and thus
no poles. A surface zero mode exists for each zero in $\det\,
\Cv$ with $|z|<1$, and the Cauchy argument theorem applied to
the contour $|z|=1$ provides a count of the total number of
zero modes at a particular surface:
\begin{align}
n_0^S (q,\Gv)  & =  \frac{1}{2 \pi i} \oint \frac{d \ln \det\, \Cv(q,z,\Gv)}{dz}
\nonumber \\
& = - \Gv \cdot \Rv_L/(2 \pi) + \tilde{n}_0^S (q,\Gv) ,
\label{Eq:count1}
\end{align}
where it is understood that $\Gv$ is the \emph{inward} normal
to the surface,  $\tilde{n}_0^S (q,\Gv)$ is the zero-mode count
for the symmetric unit cell, whose determinant has negative
powers of $z$, and $-\Gv\cdot\Rv_L$ is the local count of
Ref.~\onlinecite{KaneLub2014}. In Weyl-free regions of the
phase diagram, $\tilde{n}_0^S$ reduces to the expression, $-\Gv
\cdot \Rv_T/2 \pi$ derived in Ref.~\onlinecite{KaneLub2014}.

\begin{figure}
\includegraphics[width=.48\textwidth]{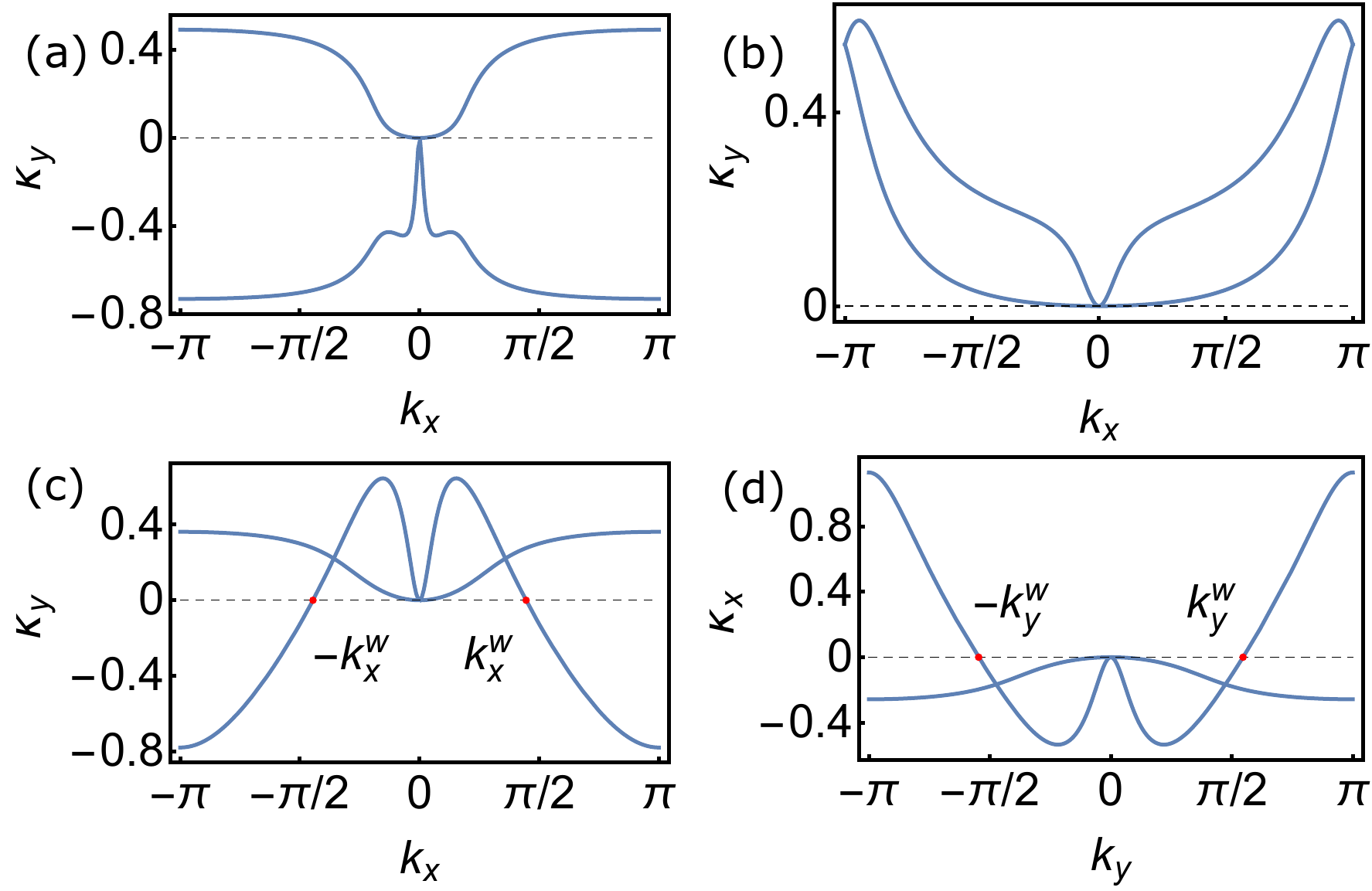}
\caption{Real part of inverse penetration depths for fully gapped
lattices with no Weyl points [(a) with $\Rv_T = (0,0)$ and (b)
with $\Rv_T = (0,1)$] and a lattice (c) and (d) with Weyl points. In (b),
 a family of zero modes has been shifted from one edge to the opposite relative
to the unpolarized case (a), while in (c) and (d) the bulk zero modes are part
of families split between two edges.}
\label{fig:inverse-decay}
\end{figure}

A Weyl point at $\kv_w \equiv (q_w,p_w)$ is characterized by an
integer winding number
\begin{equation}
n_w = \frac{1}{2 \pi i} \oint_C d \lv\cdot \nabla_{\kv} \ln \det \, \Cv
\end{equation}
where $C$ is a contour enclosing $\kv_w$. As a result, both
$n_0^S(q,\Gv)$ and $\tilde{n}_0^S (q, \Gv )$ change each time
$q$ passes through the projected position of a Weyl point.
Consider a lattice with a positive ($+1$) Weyl point at
$\kv_w^+$ and a negative ($-1$) Weyl point at $\kv_w^- = -
\kv_w^+$, and consider a surface with an inner normal $\Gv$ as
depicted in Fig.~\ref{fig:weylpath}(b) for $\Gv = (2\pi/a)
(1,1)$. The number of zero modes at $q$ is calculated from a
contour from $p=0$ to $p= G$ at position $q$. Choose $q^{1\pm}
<q_w^\pm$ and $q^{2 \pm}> q_{w}^\pm$. Because the two paths
enclose a Weyl point, the zero-mode numbers on the two sides of
the Weyl point differ by the Weyl winding number
\begin{equation}
n_0^S(q^{2\pm}, \Gv) - n_0^S(q^{1\pm}, \Gv) = n_w = \pm 1 .
\end{equation}
Thus if $n_0^S(q<q_{w}^+,\Gv) = n_1^S$, the number of zero
modes for $q_{w}^+<q<q_{w}^-$ is $n_1^S +1$, and the number for
$q>q_{w}^-$ is again $n_1^S$. \trem{If the total number of zero
surface modes at $q$ on both surfaces is $n_0^{ST}(q,\Gv)$, the
number of zero modes on the opposite surface is $
n_0^{ST}(q,\Gv) -n_1^S$ for $q< q_{w}^+$, $ n_0^{ST}(q,\Gv)
-n_1^S -1$ for $q_{w}^+<q<q_{w }^-$, and back to $
n_0^{ST}(q,\Gv) -n_1^S$ for $q>q_{w}^1$.}
Fig.~\ref{fig:inverse-decay} depicts the real part $\kappa$ of
inverse penetration lengths of surface modes with and without
Weyl points. The lengths diverge at the Weyl wavenumbers: the
surface mode turns into a bulk mode that traverses the sample.


\begin{figure}
\includegraphics{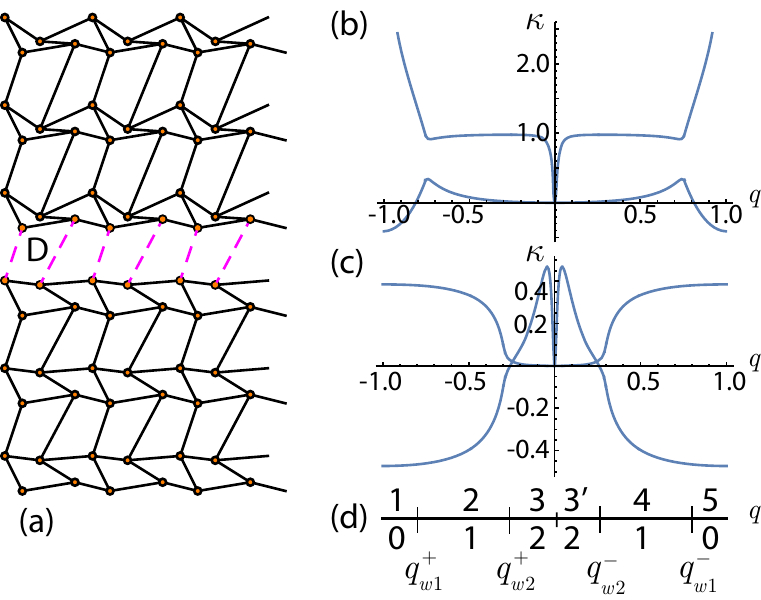}
\caption{(a) Two Weyl lattices with differently positioned Weyl points
connected at a domain wall D. (b) and (c): the inverse penetration depths of the
free surfaces of the upper and lower lattices, respectively.
The upper and lower lattices have Weyl points with
respective projections onto $q$ of $q_{w1}^{\pm}$ and $q_{w2}^{\pm}$ with $|q_{w2}^+|>
|q_{w1}^+|$. The free lower (upper) lattice has two zero modes penetrating downward (
upward) for $q_{w1}^+<q<q_{w1}^-$ ($q_{w2}^+<q<q_{w2}^-$) and one for
$|q|>|q_{w1}^+|$ ($|q|>|q_{w2}^+|$). (d) shows how the two sets of Weyl point
divide the surface BZ into five regions with $0$, $1$, $2$, $1$, and $0$
zero modes in the domain wall. The existence of bulk zero modes at $\kv=0$
divides the central region with two zero modes per $q$ into two regions.}
\label{fig:domainwall}
\end{figure}

Domain walls separating two semi-infinite lattices, which we
will refer to as the upper and lower lattices as in
Fig.~\ref{fig:domainwall}, with different topological and Weyl
characteristics harbor topologically protected zero modes. When
there are no Weyl points, the general zero-mode/SSS count is
$\nu_S^D = n_0^D - n_S^D=-\Gv\cdot (\Rv_T^L-\Rv_T^U)/2\pi$,
where $n_0^D$ and $n_S^D$ are, respectively, the number of zero
modes and SSSs in the domain wall, $\Rv_T^L$ and $\Rv_T^U$ are
the polarizations of the lower and upper lattices,
respectively, and $\Gv$ as in Eq.~(\ref{Eq:count1}) is the
\emph{inward} normal (rather than the outward normal used in
Ref.~\onlinecite{KaneLub2014}) to the top surface of the lower
lattice. When there are Weyl modes, the more general expression
in terms of winding number of the symmetric unit cell
[Eq.~(\ref{Eq:count1})] applies:
\begin{equation}
\nu^D (q,\Gv) = \tilde{n}_{0,L}^S(q,\Gv) +\tilde{n}_{0,U}^S(q,-\Gv).
\label{eq:domain-wall-count}
\end{equation}
If $\nu^D>0 (<0)$ there are zero modes (SSSs).  To prove
Eq.~(\ref{eq:domain-wall-count}) when there are no SSSs, note
that the $n_{0,L}^S+n_{0,U}^S$ combined zero modes of the free
surfaces of the upper and lower lattices each have independent
amplitudes [see Supplementary Material]. These along with the
$n_B^D$ bond stretches per unit cell associated with the bonds
required to bind the upper and lower lattices together define
an $n_B^D\times (n_{0,L}^S+n_{0,U}^S)$ domain-wall
compatibility matrix to which the Calladine-Maxwell index
theory can be applied. When the zero-mode count of the free
surfaces exceeds the number of binding bonds, the total number
of zero modes in the domain wall is $n_0^D(q,\Gv) =
n_{0,L}^S+n_{0,U}^S - n_B^D
>0$. When the topological (symmetric gauge) count is zero, there are no
domain-wall zero modes, implying that the sum of the local
counts must equal $n_B^D$. But the local counts depend only on
surface termination and do not depend on the topological count
so that $n_0^D=\tilde{n}_{0,L}^S +\tilde{n}_{0,U}^S$ is simply
the topological contribution to the zero-mode count, and
Eq.~(\ref{eq:domain-wall-count}) is recovered. Direct
calculation [see Supplementary Material] verifies that no SSSs are
introduced when binding bonds are added.
Fig.~\ref{fig:domainwall} shows how Weyl points in lattices
connected at a domain wall divide its BZ into different regions
with different numbers of zero modes.

The long-wavelength elasticity of central-force lattices is
determined by the $\kv=0$ SSSs \cite{LubenskySun2015}. In the
lattices we are considering, there are only two $\kv=0$ SSSs,
implying that there are only two positive definite eigenvalues
of the Voigt elastic matrix \cite{AshcroftMer1976}. There are
three independent strains $\stm=(\st_{xx},\st_{yy},\st_{xy})$,
and the Voigt elastic matrix must have three positive
eigenvalues and associated eigenvectors. Thus there must be one
macroscopic elastic distortion that costs no energy. This is
the Guest mode \cite{GuestHut2003} that is a feature of all
Maxwell lattices except those, like the kagome lattice, with
extra geometry-driven states of self stress
\cite{LubenskySun2015}. The elastic matrix determines the
long-wavelength dynamical matrix $\Dv(\kv) = \Qv(\kv)
\Cv(\kv)$, whose determinant is
\begin{equation}
\det \, \Dv(\kv) \propto (\st_{yy}^G k_x^2 - 2 \st_{xy}^G
k_x k_y + \st_{xx}^G k_y^2 )^2
\end{equation}
where $\st_{ij}^G$ are the components of the strain tensor of
the Guest mode. This determinant equals zero when
\begin{equation}
\frac{k_y}{k_x} = \frac{1}{\st_{yy}^G}\left[
\st_{xy}^G \pm \sqrt{- \det(\stm^G)}\right],
\end{equation}
where $\det(\stm^G) = \st_{xx}^G \st_{yy}^G - (\st_{xy}^G)^2$
is the determinant of the Guest strain matrix. Thus, to linear
order in $\kv$, there are two lines in the BZ along which there
is a zero mode provided $\det(\stm^G) <0$. These modes are
either raised to finite energy by higher order terms in $\kv$
not described by the elastic limit; or a single Weyl mode
appears along the positive and negative parts of one of the
lines. Note that this implies a quadratic rather than a linear
dispersion of phonon modes near the origin and leads to inverse
decay lengths that are proportional to $q^2$ rather than $q$ at
small $a$ as shown in Fig.~\ref{fig:inverse-decay}.

In this work, we elucidated how Weyl modes generically arise in Maxwell frames and discussed their significance
using deformed square lattices as an illustration of the more general phenomenon.
Indeed, in lattices with larger unit cells have additional phonon bands that more easily touch, generically leading to  Weyl points and even multiple pairs thereof [see Supplementary Material].
Thus, our conclusions can be
readily extended to other Maxwell lattices like origami
metamaterials~\cite{Chen2015}, random spring networks and
jammed sphere packings~\cite{SussmanLub2015}, and 3D distorted
pyrochlore lattices~\cite{StenullLub2015a} that fulfill the
Maxwell condition. We also expect the presence of Weyl modes to
impact the nonlinear response (e.g.\ buckling) in the bulk as
demonstrated for edge
modes~\cite{PauloseVit2015-b,ChenVit2014}.

We are grateful to Charles Kane for many informative
discussions and suggestions. (DZR) thanks NWO and the Delta Institute of Theoretical Physics for supporting his stay at the Institute Lorentz.
MJF was supported by the Department of Defense (DoD) through the National Defense Science \& Engineering Graduate Fellowship (NDSEG) Program.
This work was supported in part by
DMR-1104707 and DMR-1120901 (TCL), as well as FOM and NWO (BGC,
VV).
\bibliography{RPP}

\appendix

\section*{Supplementary Material}

\subsection*{Topological polarization via winding number}
\label{sec:zeroandwinding}

In this section, we calculate the topological polarization of
lattices with and without Weyl modes.
We consider lattices with particles lying at sites $\basisvec_i + \cellvec$, where $\{\basisvec_i\}$ are the basis vectors giving the site positions within a unit cell and $\cellvec \equiv n_1 \latvec_1 + n_2 \latvec_2$ is the position of the cell, with $\latvec_1, \latvec_2$ the Bravais vectors. 
 A phonon mode
may be expressed as 
$\uvec(\cellvec)=\exp
 \left( i \momvec  \cdot \cellvec\right)
 \uvec$
, where $\uvec$ is a vector of displacements
within a crystal cell. More generally, we can locally satisfy the dynamics with
modes that exponentially grow and decay as well as oscillating, taking the form:

\begin{align}
\label{eq:uvecform}
\uvec(\cellvec)=\exp
 \left[\left( i \momvec - \magvec \right) \cdot \cellvec\right]
 \uvec \equiv \zone^{n_1} \ztwo^{n_2} \uvec,
\end{align}

\noindent where the components of $\magvec$ describe the mode's growth and decay and are plotted in Fig.~\ref{fig:inverse-decay} of the main text. Note that $\zone, \ztwo$ are given by

\begin{align}
z_{1} = \exp \left[\left(i \mathbf{k} - \mathbf{\kappa}\right)\cdot \latvec_1 \right],
\\ \nonumber 
z_{2} = \exp \left[\left(i \mathbf{k} - \mathbf{\kappa}\right)\cdot \latvec_2 \right].
\end{align}

\noindent
If $\magvec = 0$, the complex numbers $\zone, \ztwo$ have unit
magnitude and
represent a purely oscillatory phonon mode in the bulk.

 For a deformed square lattice (choosing a symmetric
unit cell) the four intercellular bonds then generate terms in
the compatibility matrix proportionate to $\zone, \ztwo,
\zone^{-1}, \ztwo^{-1}$ so that for fixed $\ztwo$  the
determinant  has the form

\begin{align}
\det(\Cv(\zone,\ztwo)) \sim  b_{-1}(\ztwo) \zone^{-1} + b_0(\ztwo) + b_1(\ztwo) \zone,
\end{align}

\noindent where the $\{ b_i \}$ are complex functions of
$\ztwo$. From this, we could solve for the two exact values of
$\zone$ where the determinant vanishes, signaling the presence
of a zero mode. Of particular significance is whether $|\zone|$
is greater than one, indicating a zero mode on the right edge
of the system, or less than one, indicating a zero mode on the
left edge. 
This can be determined entirely in terms of the
winding of the bulk modes around the Brillouin zone, those in
which $\zone$ has the form $\exp(i \momv2_x)$, for a winding number

\begin{align}
\frac{1}{2\pi}
&\int_{0}^{2 \pi} d \momv2_x 
\ln \frac{\partial}{\partial \momv2_x} \det \left[ \Cv(\exp(i \momv2_x),\exp(i \momv2_y))\right]
\\ \nonumber
=
\frac{1}{2 \pi}
&\oint_{C} d \zone \frac{\frac{\partial}{\partial \zone} \det(\Cv(\zone,\exp(i \momv2_y)))
}{\det(\Cv(\zone,\exp(i \momv2_y)))}.
\end{align}

\noindent Because $\momv2_x = 0$ and $\momv2_x = 2\pi$
correspond to the same point in the Brillouin zone, the real
parts of these integrals vanish and the imaginary part is the
change in phase of $\det(\zone, \exp(i \momv2_y))$ as one
follows the contour $C$ around the unit circle. According to
the Argument Principle, obtained via performing a contour
integral, this integral is $2 \pi i (N - P)$, where $N$ and $P$
are respectively the number of zeroes and poles of $\det(\zone,
\exp(i \momv2_y))$ enclosed by the contour. Thus, windings of
$-2\pi$, $0$, and $2 \pi$ correspond respectively to 0, 1 and 2
zeroes of $\det(\zone, \exp(i \momv2_y))$ lying in the unit
circle and consequently 0, 1 or 2 zero modes on the left edge
of the system, with the remaining 2, 1, or 0 lying on the
right.

To obtain all the zero modes we must repeat this winding
calculation for all values of $\momv2_y$, but for lattices
without Weyl modes this proves trivial---the path may be
continuously deformed across the Brillouin zone (the zero at
the origin has zero winding number and so is trivial) and so we
find that none, half, or all of the horizontal zero modes lie
on the left wall of the system. Similarly, calculating the
winding of $\det(\exp(i \momv2_x), \exp(i \momv2_y))$ as
$\momv2_y$ advances by $2 \pi$ determines how many zero modes
lie on the top and bottom edges of the system. Combining these
two values into a vector gives us the \emph{topological
polarization} of a lattice, which points in the average direction
of zero modes:

\begin{align}
\toppol=\frac{-1}{2\pi} \left( \Delta \phase(\momvec \rightarrow \momvec + 2 \pi (1,0), \Delta \phase(\momvec \rightarrow \momvec + 2 \pi (0,1)\right).
\end{align}

\noindent
This topological polarization is uniquely defined and independent of $\kv$ provided that there are no Weyl modes.  $\toppol$, along with the local vector $\Rv_L$, determines the number of zero modes on a surface specified by a reciprocal lattice vector $\Gv$ as discussed in Eq.~(\ref{Eq:count1}).
Consider, as a simple example, an $L
\times L$ lattice where the open boundary is made up of straight cuts along reciprocal lattice vectors that do not split any cells.
For such a lattice with topological
polarization $(0,0)$ there are $L$ zero modes on each edge, while a lattice
with topological polarization $(0,1)$ has $2 L$ modes on its top edge
and none on its bottom edge.

\subsection*{Winding of the phase across the Brillouin zone and about a Weyl point}
\label{sec:winding}

\begin{figure}[hh]
\includegraphics[width=.4\textwidth]{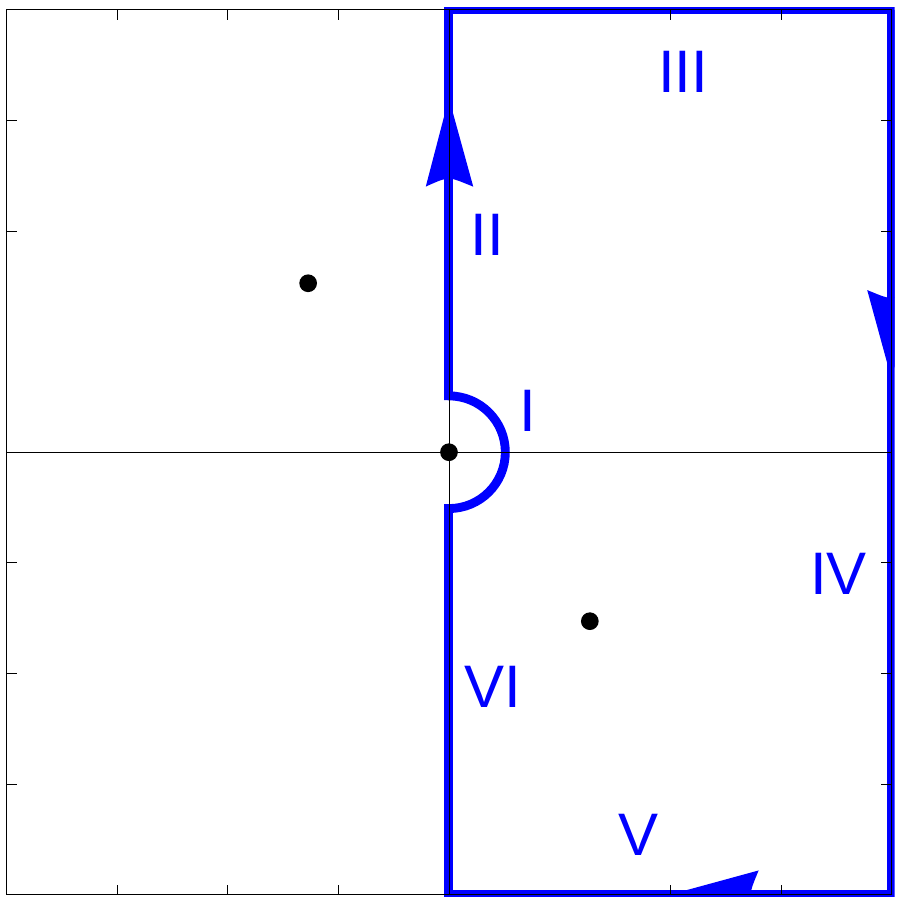}
\caption{The path around the Brillouin zone along which the winding is evaluated. A winding of $+1$ or $-1$ indicates the presence of a Weyl point (shown as a black dot), while a winding of $0$ indicates that the lattice has no Weyl modes. The indentation along path I avoids the zero modes at $\momvec=0$. When no Weyl points are present, the winding along the path VI, I, II yields one component of the topological polarization, with the other component being given by a similar horizontal path, as described in the main text.}
\label{fig:contour}
\end{figure}

As discussed in the previous section, the winding of the phase
of $\det \, \Cv(\momvec)$ through the Brillouin zone determines
both the presence of Weyl modes and the topological
polarization of the lattice. Consider the path shown in
Fig.~\ref{fig:contour}. We consider only noncritical lattices,
so that the determinant vanishes only at $\momvec = (0,0)$ and
possibly at a pair of Weyl points located at $\pm \momvec_W$.
The path chosen thus has a well-defined phase and encloses at
most one Weyl point. A winding of $\pm 1$ indicates the
presence of a Weyl point, while a winding of $0$ indicates a
lattice without finite-wavenumber zero modes.

Consider first the segments II and VI. From the form of the
determinant, which must have a double root at $\zone = \ztwo =
1$ corresponding to the two translational modes, it may be
shown that along either of these segments the determinant is
real and nonzero, so its winding does not change.

Note that $\Cv$ is periodic under $\momvec \rightarrow \momvec
+ 2 \pi (0,1)$. Thus, while the winding along segment III may
be finite, it must be canceled by the return path along segment
V. Thus, the winding comes entirely from segment I (the
``origin term'') and segment IV (the ``edge term'').

Along segment IV, $\momv2_x = \pi$ and

\begin{align}
\label{eq:segfour}
\det \, \Cv(\momvec) \sim a \exp \left( i \momv2_y \right) + b + c \exp \left( -i \momv2_y \right),
\end{align}

\noindent where $(a,b,c)$ are real lattice-dependent numbers.
As $\momv2_y$ goes from $\pi$ to $-\pi$, $\det \, \Cv(\momvec)$
travels along an elliptical path that may be either clockwise
or counterclockwise and may or may not enclose the origin,
leading to a winding

\begin{align}
\textrm{winding}(IV) = \textrm{Sign}(c-a)
\\ \nonumber
\times \left[ \textrm{Sign}(a + b + c) - \textrm{Sign}(b-a-c)\right]/2.
\end{align}

Along segment I, $\det \, \Cv(\momvec)$ takes its long-wavelength
form

\begin{align}
\det \, \Cv(\momvec) \sim a \momv2_y^2 + b \momv2_x \momv2_y + c \momv2_x^2 +
\\ \nonumber
i \left(d \momv2_x \momv2_y^2 + e \momv2_x^2 \momv2_y\right) + O(|\momvec|^4),
\end{align}

\noindent where these lattice-dependent real parameters
$(a,b,c,d,e)$ differ from those of Eq.~(\ref{eq:segfour}). In
the neighborhood of the origin, the determinant has a real part
$O(|\momvec|^2)$ and an imaginary part $O(|\momvec|^3)$. Thus,
its phase can only appreciably change when the real part passes
through zero, along lines

\begin{align}
\momv2_y = \alpha_{\pm} \momv2_x; \quad \alpha_{\pm}  = \left(-b \pm \sqrt{b^2-4 a c}\right)/(2 a).
\end{align}

If the discriminant is negative, segment I does not pass
through any such lines, and so the determinant does not undergo
any sort of winding. Otherwise, it passes by the origin twice,
with imaginary parts

\begin{align}
v_1 = \left(d \alpha_1+ e \alpha_1^2\right)|\momvec|^3,\\
v_2 = \left(d \alpha_2+ e \alpha_2^2\right)|\momvec|^3,
\end{align}

\noindent where $\alpha_1$ is $\textrm{Min}(\alpha_\pm)$, so
the segment passes through $\momv2_y = \alpha_{1} \momv2_x$
first. Clearly,  segment I begins and ends near $a k_y^2$ and
undergoes finite winding only when $v_1$ and $v_2$ differ in
sign. The winding is

\begin{align}
\textrm{winding}(I) = \textrm{Sign}(a)\left[ \textrm{Sign}(v_1)- \textrm{Sign}(v_2) \right]/2.
\end{align}

Thus, the sum of the origin term and the edge term gives the
winding around half of the Brillouin zone, which indicates the
presence or absence of Weyl modes in the lattice. In the
absence of such Weyl modes, the winding along segments VI, I
and II (of which only the second is nonzero) is the winding
along any path that increases by $(0,2\pi)$. This is one
component of the topological polarization. The other component
is readily obtained by repeating the analysis along a
similarly-indented path from $(-\pi,0)$ to $(\pi,0)$. Thus, the
same analysis of winding numbers that identifies lattices with
Weyl modes determines the topological polarizations of those
without such modes.

For lattices with larger unit cells, the calculation becomes more complicated. Each intercellular bond introduces an additional factor of $\zone, \ztwo,
\zone^{-1}$ or  $\ztwo^{-1}$ into $\det \, \Cv$, increasing the polynomial degree. Weyl points are zeroes of this expression satisfying 
$|\zone|=|\ztwo|=1$. Hence, in larger unit cells allow more solutions of these equations and hence more Weyl points, or even multiple pairs of Weyl points.

\subsection*{Domain-wall zero modes: count and structure}

As discussed in the text following
Eq.~(\ref{eq:domain-wall-count}), the number of zero modes and
their structure can be obtained from the eigenvalues for $z$
and the associated eigenvectors of the zero modes of the free
surfaces of the upper and lower lattices that are joined
together at the wall. Here we outline the derivation of this
result. The upper and lower lattices have, respectively,
$n_{0,U}^S$ and $n_{0,L}^S$ zero surface modes modes at their
free surfaces. Let the $z$-eigenvalues and associated
eigenvectors for zero modes of the upper and lower lattices at
given $q$ along the wall be $z_{\mu +}(q)$, $(\av_1^{\mu
+},\av_2^{\mu +},\av_2^{\mu +},\av_2^{\mu +})$ and $z_{\mu
-}(q)$, $(\av_1^{\mu -},\av_2^{\mu -},\av_2^{\mu -},\av_2^{\mu
-})$, respectively, where $\av_n^{\mu {\pm}}$ is the
displacement of site $n$ in the unit cell and by definition
$\Cv_{\pm}(q,z_{\mu\pm},G) \av_{\mu\pm}=0$ where $\Cv_{\pm}$ is
the compatibility matrix of the upper ($+$) or lower ($-$)
lattice. Any linear combination of zero modes,
\begin{equation}
\uv_n^{\pm} (q,n_x,n_y) = \sum_{\mu_{\pm}} A^{\mu \pm} \av_n^{\mu \pm}
(z_{\mu {\pm}}(q))^n_y e^{i n_x q}
\label{eq:uv-zero}
\end{equation}
is also a zero mode, where $n_x$ and $n_y$ are the positions of
unit cell along $x$ and $y$.  Denote the $n_B^D$ unit vectors
along the bonds that bind the upper an lower lattices by
$\bv_{\alpha}$. Each bond $\alpha$ connects an upper lattice
boundary site $n_+(\alpha)$ to a lower lattice site
$n_-(\alpha)$.  The lengths of these binding bonds must not
change in a domain-wall zero mode.  If we restrict
displacements of the upper and lower lattices to their surface
modes, the length of no bond will changes provided, the binding
bonds satisfy
\begin{equation}
\bv_{\alpha} \cdot [\uv_{n_+(\alpha)}^+ (q,n_x,0)- \uv_{n_-(\alpha)}^- (q,n_x,0)]=0
\label{eq:balpha})
\end{equation}
for all $\alpha$ and any zero-mode displacements $\uv_{n_\pm
(\alpha)}^\pm (q,n_x,0)$ of the upper and lower lattices. Then,
Eqs.~(\ref{eq:uv-zero}) and (\ref{eq:balpha}) can be written as
$\Cv^D \cdot \Av = 0$, where $\Av$ with components
$A^{\mu_\sigma}$, $\sigma = \pm$,  is the $(n_{0,U}^S +
n_{0,L}^S)$-dimensional column vector of displacement
amplitudes and the components of the $n_B^D\times (n_{0,U}^S +
n_{0,L}^S)$ domain-wall compatibility matrix are
\begin{equation}
C_{\alpha, \mu \sigma}^D= \sigma \bv_{\alpha} \cdot \av_{n_{\sigma}(\alpha)}^{\mu {\sigma}} .
\end{equation}
The null space of this matrix is the space of zero modes of the
domain wall.

\begin{figure}
\includegraphics{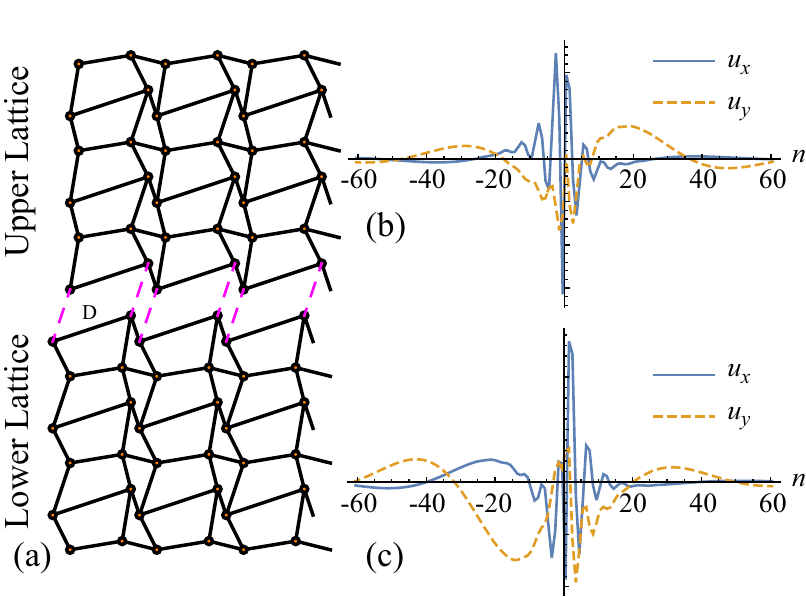}
\caption{(a) A lattice with a domain wall connecting a lower lattice to an upper lattice,
which is a 180 degree rotation of the lower lattice. (b) and (c) Spatial representations of
the displacement of site 1 of the upper lattice (positive cell index $n$) and site 2 of
the lower lattice (negative $n$) (following the naming convention of Fig.~\ref{fig:weylpath})
for the two zero modes at $q=2.0$.}
\label{fig:domain-wall-2}
\end{figure}

\begin{figure}[H!]
\includegraphics{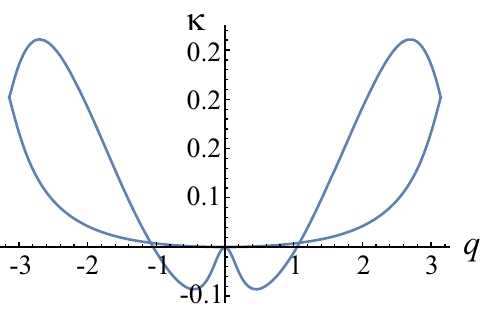}
\caption{The real part of the inverse penetration depth $\kappa$ as a function of wavenumber
$q$ for a surface parallel to the $x$-axis.  A positive value of $\kappa$ implies a decaying
solution into the bulk.  Thus, there are two decaying solutions for $q=2.0$.}
\label{fig:kappas}
\end{figure}

As a specific example, consider the lattice shown in
Fig.~\ref{fig:domain-wall-2} (a) with a domain wall connecting
a lattice with its inverted image.  The  real parts of the
inverse decay lengths of these two lattices are the same and
are depicted in Fig.~\ref{fig:kappas}.  For $q>1$, there are
two positive decay lengths yielding $n_{0,2}^S = n_{0,U}^S =
2$.  Two bonds per unit cell connect the two lattices, so
$n_B^D = 2$ and $\Cv^D$ is a $2 \times 4$ dimensional matrix
\begin{equation}
\Cv^D =
\begin{pmatrix}
\bv_1\cdot \av_4^{1 +} & \bv_1\cdot \av_4^{2 +} & -\bv_1\cdot \av_3^{1 -} &  -\bv_1\cdot \av_3^{2 -} \\
\bv_2\cdot \av_1^{1+} & \bv_2\cdot \av_1^{2+} & -\bv_2\cdot \av_2^{1-} &  -\bv_2\cdot \av_2^{2 -}
\end{pmatrix}
\end{equation}
As expected, this matrix has a two-dimensional null space, and
there are two zero domain-wall zero modes.  Their wave
functions for $q=2.0$ are depicted in
Fig.~\ref{fig:domain-wall-2} (b) and (c).

\end{document}